\documentclass{article}     
\usepackage{graphicx}
\usepackage{epstopdf}
\title{On the proton radius problem}  
\author{M.M. Giannini\\
Dipartimento di Fisica dell'Universit\`a di Genova\\
and \\
I.N.F.N., Sezione di Genova\\
\\
E. Santopinto\\
I.N.F.N., Sezione di Genova\\
}

\date{}
\begin{document}             

\maketitle         

\begin{abstract}
The recent values of the proton charge  radius obtained by means of muonic-hydrogen laser spectroscopy are about $4\%$ different from the electron scattering data. It has been suggested that  the proton radius is actually measured in different frames  and that, starting from a non relativistic quark model calculation, the Lorentz transformation of the form factors accounts properly for the discepancy. We shall show that the relation between the charge radii measured in different frames can be determined on very general grounds by taking into account the Lorentz transformation of the charge density. In this way, the discrepancy between muonic-hydrogen laser spectroscopy and electron scattering data can be removed.
\end{abstract}

An accurate knowledge of the charge radius of the proton  is  fundamental for a deep understanding of its internal structure.  The average value, recommended by the 2010-CODATA review \cite{codata} is 0.8775(51) fm, obtained from hydrogen atom and electron-proton elastic scattering measurements. Such value has been reinforced by the recent results of the experiments performed at Jefferson Lab \cite{zhan} and MAMI \cite{bern}

Therefore it came as a surprise that in the recent measurements by means of muonic-hydrogen laser spectroscopy \cite{pohl,ant} the proton charge radius turns out to be about $4\%$ different. In particular, the latest result \cite{ant} 0.84087(39) fm is at $7 \sigma$ variance with respect to the CODATA value.

The various methods for measuring the proton radius have been discussed and their results compared in \cite{mill}. An analysis of how the present average value has been stabilized shows that the electron scattering value seems to converge definitely to 0.8775(51) fm \cite{codata}, while part of the hydrogen atom measurements, much less precise that the muonic ones, support also lower values of the charge radius. There seems to be no realistic way of explaining such a discrepancy and therefore we are in presence of a real puzzle \cite{mill}.

An interesting attempt to solve such a problem has been done in ref. \cite{rob}. The  idea is that in electron scattering and muonic hydrogen atom one actually measures the same quantity in two different reference frames. In fact, in the case of electon scattering one refers to the Breit system, while in the hydrogen atom one can think at the proton as being at rest. Therefore the issue of how form factors are modified applying a Lorentz trasfomation is addressed using the calculations in ref. \cite{LP}. Their starting point is    the calculation of the charge form factor in a non-relativistic quark model, which is typically performed in the proton rest frame. However, in order to compare the results with the electron scattering data, one has to boost to the Breit frame, where the initial and final momenta of the proton are $\vec{p}$ and $-\vec{p}$, respectively, the virtual photon momentum being then $2 \vec{q}$. The effect of the Lorentz transformation can be taken into account  by means of the following relation \cite{LP}
\begin{equation}
F_B(\vec{q})~ = ~\tau(\vec{q}~^2)^{\frac{1-n}{2}}  ~      F_{RF}(\frac{\vec{q}~^2}{\tau})
\label{LP}
\end{equation}
where $ F_{RF}$ and  $F_B$ are the charge from factors in the rest and Breit frames, respectively. In Eq. (\ref{LP}) n is the number of costituent quarks in the proton and
\begin{equation}
\tau~ = ~1      ~ + ~     \frac{\vec{q}~^2}{4 M^2}
\label{Lo}
\end{equation}
where $M$ is the proton mass . Taking n=3 and defining the root mean square charge radius as
\begin{equation}
<r^2>~ = ~- 6 \frac{d F(\vec{q}~^2)}{d \vec{q}~^2} |_{\vec{q}~^2=0}
\label{rms}
\end{equation}
then, from Eq.( \ref{Lo}) it is easy to obtain the relation \cite{rob}
\begin{equation}
<r^2>_B~ = ~<r^2>_{RF}~ + ~\frac{3}{4 M^2}
\label{rob}
\end{equation}
therefore, adding the boost term to the rest frame value the discrepancy is almost completely removed \cite{rob}.

This result is certainly very interesting, however it can be criticized for two fundamental reasons. The first one is that the elastic electromagnetic form factors of the nucleon are Lorentz invariant and do not transform while going from one frame to another one \cite{inv}. The second reason is that the result is based on a quark model calculation. The consideration of a particular quark structure of the proton, although  widely accepted, introduces  in the calculation an unacceptable model dependence. 

Actually, the relation of Eq. (\ref{LP}) should be more correctly interpreted as the relativistic correction of a non relativistic calculation. In this sense, there have been other  approaches to the relativistic corrections, leading to results quite different from Eq. (\ref{rob}). In fact, for example, in ref. \cite{Holz96}, the relativistic corrections to the non relativistic soliton calculation of the proton charge form factor lead to the relation
\begin{equation}
F_B(Q^2)~ = ~  F_{RF}(\frac{Q^2}{\tau})
\label{holz}
\end{equation}
where $Q^2 = -q^2$, with $q^2$ being the virtual photon tetramomentum square;  in the Breit frame one has $Q^2=\vec{q}~^2$. From Eq. (\ref{holz}) one obtains
\begin{equation}
<r^2>_B~ = ~<r^2>_{RF}
\end{equation}

Furthermore, in ref. \cite{mds} the relativistic corrections, applied to the elastic form factors obtained with the non relativistic hypercentral Constituent Quark Model \cite{pl}, lead to
\begin{equation}
F_B(Q^2)~ = ~A ~        F_{RF}(Q^2 \frac{M^2}{E^2})
\label{h}
\end{equation}
where $E$ is the nucleon energy in the Breit frame and A a kinematical factor (the explicit expression can be found in \cite{mds}). From Eq. (\ref{h}), one gets \cite{mds}
\begin{equation}
<r^2>_B~ = ~<r^2>_{RF}~ + ~\frac{6}{M^2}
\end{equation}

Therefore, we have in this way shown that using quark models for the determination of the influence of the reference frame on the charge radius a strong model dependence is introduced. Moreover, it is hard to understand why one should refer expicitly to the nucleon internal structure in a discussion concerning relativistic properties.

 In the following we shall show that Eq.(\ref{rob}) can be obtained from very general arguments.

The relation between the charge form factor and r.m.s. radius is obtained from the Fourier transform \begin{equation}
F(\vec{q}~^2)~ = ~ \int d^3r ~ e^{i\vec{q} \cdot \vec{r}} \rho(\vec{r})
\label{FF}
\end{equation}
where  $\rho(\vec{r})$ is the charge density. Performing an expansion for small values of $|\vec{q}|$ and assuming that the charge density is symmetric for spatial reflections, one gets
\begin{equation}
F(\vec{q}~^2)~ \simeq ~ 1 ~ - ~ \frac{1}{6}r ~ <r^2> ~|\vec{q}|^2
\end{equation}
or equivalently Eq.(\ref{rms}) reported above.

The charge density depends on the reference frame and $\vec{q}$ should be considered as the virtual photon tri-momentum in the same frame. Then also the Fourier transform of the charge density is frame dependent and should not be identified with a form factor, say the Sachs form factor $G_E(Q^2)$, which is a Lorentz invariant. The point is that the Fourier transform in Eq.(\ref{FF}) can be expressed in terms of Lorentz invariant form factors in a way that depends on the reference frame.

To see this let us start from the general form of the nucleon electromagnetic current in momentum space:
\begin{equation}
J^ \mu(\vec{p}_f,\vec{p}_i)~=~\bar{u}(\vec{p}_f) [F_1 \gamma^{\mu} +  F_2  \frac{i}{2M} 
\sigma^{\mu \nu} q_{\nu}] u(\vec{p}_i)
\end{equation}
where $F_1=F_1(q^2) $ and $F_2=F_2(q^2) $ are the Dirac and Pauli form factors, respectively, $q=p_f-p_i$ is the virtual photon tetramomentum and $u(\vec{p}_f)$, $u(\vec{p}_i)$ are spinors obeying the Dirac equation. This form is obtained imposing that the current should be covariant and conserved and that the nucleon is on the mass-shell \cite{inv}. In particular, covariance implies that the form factors $F_1$ and $F_2 $ must be Lorentz invariant, otherwise the current should not transform as a tetravector.

An alternative and equivalent expression is obtained using the Gordon expansion $\gamma^{\mu} \gamma^{\nu} ~=~  g^{\mu \nu} - i \sigma^{\mu \nu}$:
\begin{equation}
J^ \mu(\vec{p}_f,\vec{p}_i)~=~\bar{u}(\vec{p}_f) [(F_1 ~ + ~ F_2)\gamma^{\mu} -  F_2  \frac{p_i^{\mu}  + p_f^{\mu}}{2M}] u(\vec{p}_i)
\end{equation}

Using the explicit form of the Dirac spinors (with the normalization $\bar{u} u=1$) and considering the Breit frame (where, as already mentioned,  $\vec{p}_i=- \vec{p}_f= -2 \vec{q}$) one obtains the following expression for the charge density in momentum space:
\begin{equation}
\rho_B(Q^2)~ = ~ J^ 0(-\vec{p},\vec{p})~=~F_1(Q^2) ~ - ~ \frac{Q^2}{4 M^2} ~  F_2(Q^2)~ = ~G_E(Q^2)
\end{equation}
where $G_E(Q^2)$ is the Sachs form factor \cite{sachs}. 

We should note that in the Breit frame $Q^2=-q^2$ is simply given by $|\vec{q}|^2$, therefore we can write
\begin{equation}
<r^2>_B~ = ~-6 \frac{d \rho_B(|\vec{q}|^2)}{d |\vec{q}|^2} |_{|\vec{q}|^2=0}~ = ~- 6 \frac{d G_E(Q^2)}{d Q^2} |_{Q^2=0}
\end{equation}

A similar calculation can be performed in the nucleon rest frame, obtaining
\begin{equation}
\rho_{RF}(k^2)~=  J^ 0(\vec{k},0)=\sqrt{\frac{E_k+M}{2M}}(
F_1 + F_2  -  \frac{E_k+M}{2 M}   F_2)=\sqrt{1+\tau} G_E(Q^2)
\end{equation}
where $\vec{k}$ is the photon tri-momentum in the nucleon rest frame and $E_k$ is the final nucleon energy. In this frame we have $Q^2=2M(E_k-M)$. With the definition
\begin{equation}
<r^2>_{RF}~ = ~- 6 \frac{\rho_{RF}(k^2)}{d k^2} |_{k^2=0}
\end{equation}
one gets
\begin{equation}
<r^2>_{RF}~ = ~<r^2>_{B}~ - ~\frac{3}{4 M^2}
\label{gs}
\end{equation}
which is precisely Eq.(\ref{rob}).

This means that, in order to obtain the charge radius in any given reference frame, one must use  the charge density and the trimomentum both evaluated in the same frame. As already stated above, the charge density in momentum space is not a Lorentz invariant and its expression in terms of the Lorentz invariant form factors $F_1$ and $F_2$ depends on the reference frame.

In conclusion, we can say that the charge radius obtained from electron scattering measurements should be identified with $<r^2>_{B}$, since it is related to the Sachs form factor $G_E(Q^2)$. As for the muonic hydrogen atom experiments, the proton can be fairly considered in its rest frame and Eq. (\ref{gs}) explains the major part of the difference in the measured r.m.s. radii. The problem remains with the electron hydrogen atom experiments, because the resulting proton radius seems to be more similar to the electron scattering data.

 In any case, we think that the transformation properties derived above should be taken properly into account when adressing the proton radius puzzle.
\\


\end{document}